# Laser array of coherent beam combination system revisited: angular domain perspective and fractal-based optimization

Tianyue Hou [1], Qi Chang [1], Pengfei Ma [1]*, Jinhu Long [1] and Pu Zhou [1]*

[1] College of Advanced Interdisciplinary Studies, National University of Defense Technology, Changsha 410073, China

**Abstract**: Coherent beam combination (CBC) of fiber lasers holds promise for achieving high brightness laser systems, which have given rise to widespread applications such as particle accelerator, space debris removal, and industrial fabrication. The emitting laser array of CBC systems offers intriguing features in terms of agile beam steering, flexible beam shaping, and high scalability for output power and array elements. However, the theoretical model of the laser array in CBC systems is less well explored beyond the routine angular-spectrum method, where methods for optimizing the laser array configuration are more limited. Here, we explore the theory for the laser array of CBC systems in the view of angular domain. The laser array is represented by the composition of angular harmonics, the orthogonal basis over the azimuthal plane, and we elucidate the formation of mainlobe and sidelobes of the far-field interference pattern by using the orbital angular momentum spectrum analysis and azimuthal decomposition. Based on our findings, a fractal-based laser array configuration is proposed to enhance the performance of the combining system. Our work offers a deeper insight into the theoretical study and application of laser beam combination and opens opportunities for the further optimization of CBC implementations.
**Keywords**: coherent beam combination; laser array; angular domain; orbital angular momentum

## 1. Introduction

There has been a continuing interest to scale the beam power and brightness with good beam quality ever since the laser's invention. To remove the power barrier of single lasers, the development of laser beam combination with high efficiency has been an ongoing effort [1]. Coherent beam combination (CBC) of fiber lasers, as a promising way to achieve high brightness laser systems, can enhance the output power from laser sources while maintain the advantages of fiber lasers including the good beam quality, compact structure, and high efficiency [2,3]. In the CBC system, relative phases of the laser array elements are controlled to form constructive interference in the far-field. With the significant progress in this century, CBC technique has been employed in versatile scientific frontiers, accordingly spurring a wide range of applications such as the next-generation particle accelerator and nuclear waste transmutation [4,5], space debris tracking and removal [6], and advanced laser architectures for spatial light structuring [7-9] and industrial processing [10]. Generally, a typical CBC system mainly consists of three sections, namely the high-power combined elements, dynamic phase control system, and emitting laser array configuration. Among them, the performance enhancement of combined elements relies on the extended physical knowledge of the comprehensive nonlinear and thermal effects in the active fiber, and the output power of a single laser beam in the CBC system has been remarkably scaled to multiple kilowatts level [11-13]. Similarly, the phase control system has been developed by deepening the understanding of the optical field information extraction methods and algorithms [14-19], and accordingly, efficient phase locking of more than one hundred channels fiber lasers has been recently realized [20].

Emitting laser array configuration, as another vital section of the CBC system, significantly influences the beam quality, far-field energy distribution, and combining efficiency of the system [21]. Routinely, the celebrated angular-spectrum method in the theoretical framework of Fourier optics is utilized to study the emitting laser array configuration [22]. Representing the laser array by the composition of tilted plane waves with different spatial frequency, the optical field of the coherently combined beam in the far-field can be calculated, which typically contains a central mainlobe of concentrated energy and surrounding sidelobes that limit the combining efficiency [1,23]. Despite the salient utility for the calculation of optical field distributions, the present theoretical model of emitting laser array configuration is difficult to elucidate the underlying principle for the formation of the mainlobe and sidelobes, and therefore, sidelobes suppression for CBC systems is always intuitively recognized as a long-standing *engineering* problem [1,2,21,24]. There is now increasing interest in exploring the theory of laser arrays in both classical and quantum regimes, and its important implications have been confirmed in developing a variety of advanced optics and photonic systems [25-28]. To fulfill the continuing desire in the enhancement of combining efficiency, it is thus essential to revisit the principle of the laser array in CBC systems with a deeper insight and further extend the optimization of the system design to a new perspective.

In this paper, we extend the theory for the emitting laser array configuration of CBC systems to the view of angular domain. Utilizing the angular harmonics in form of $\exp(il\theta_0)$ to decompose the optical field, we represent the laser array of the system by the superposition of the orthogonal basis over the azimuthal plane. Based on the angular harmonics representation of laser array, the investigations of orbital angular momentum (OAM) spectrum and azimuthal decomposition are carried out, and the OAM modes of the laser array that contribute to the formation of mainlobe and sidelobes in the far-field are discussed in detail. Moreover, an optimization method for system design is proposed by incorporating the symmetric fractal into our findings in the angular domain perspective of emitting laser array configuration. We demonstrate that the optimized laser array with a self-similar fractal structure can efficiently suppress the sidelobes and improve performance of the system. This work could be useful for the theoretical study and practical implementation of laser beam combination.

## 2. Principle

The typical architecture for the CBC implementation is schematically shown in Fig. 1. The seed laser (SL) output is sent through a pre-amplifier (PA) to preliminarily scale the power. Then, the output of PA is split into multiple channels by a fiber splitter (FS) and sent to phase modulators (PMs) used for phase control. The output laser beam of each PM is used as the input of cascaded fiber amplifiers (CFAs), which amplify the laser beam to high power. The high-power outputs from the CFAs are tiled in side-by-side arrangement and collimated by the collimator array at the source plane. The output laser beams of the collimator array are sampled by a high reflective mirror (HRM1) in free space. The reflected beams are the high-power output of the CBC system, while the transmitted part of low power subsequently passes through the HRM2 and a focus lens (L). The focused laser beams are sampled by a beam splitter (BS) and transformed to the far-field. Part of the focused beams are captured by a CCD positioned at the focal plane, thus the far-field intensity profile of the coherently combined beam can be observed. The other part is coupled into a photodiode detector (PD), which sends the signals that contain the optical field information to the field-programmable gate array (FPGA) controller. By processing the received signals and operating the phase control algorithm, the FPGA controller applies the phase control signals to the PMs to realize the real-time compensation of dynamic phase noises in the system. When the system is in closed-loop, the output laser beams from the collimator array are phase-locked, and accordingly, constructive interference would be formed in the far-field.

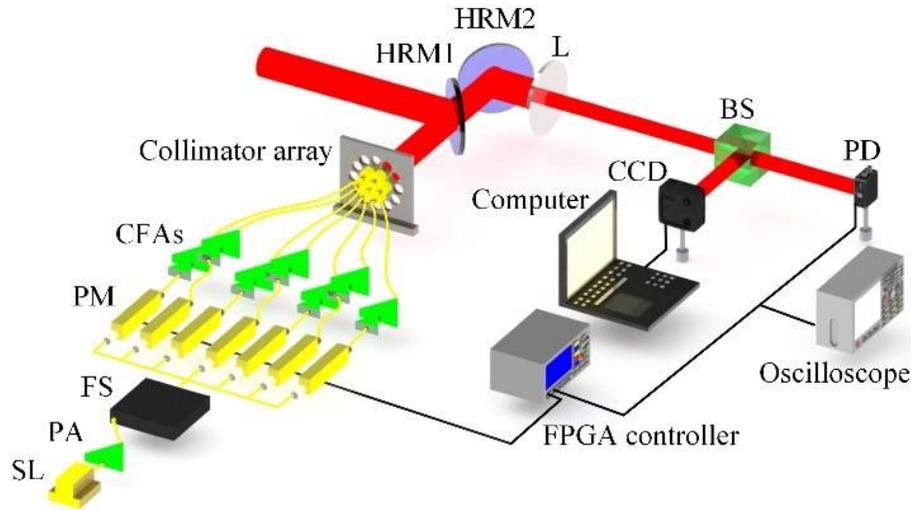

Fig. 1. Schematic of CBC architecture. (SL: seed laser; PA: pre-amplifier; FS: fiber splitter; PM: phase modulator; CFAs: cascaded fiber amplifiers; HRM: high reflective mirror; L: focus lens; BS: beam splitter, PD: photodiode detector.)

To investigate the optical field of the CBC system, the theoretical model of emitting laser array at the source plane requires to be constructed, and the optical field distribution of the coherently combined beam in the far-field can be correspondingly calculated. Conventionally, constructing the theoretical model of the emitting laser array is treated in the *Cartesian* domain. Consider a coherent laser array consists of $N$ beamlets, which are linearly polarized fundamental Gaussian modes truncated by circular apertures with diameter of $d$. The amplitude, waist width, and wavelength of each beamlet are $A_0$, $w_0$, and $\lambda$, respectively. The optical field of the laser array at the source plane is given by

$$E(x_0, y_0) = A_0 \sum_{j=1}^{N} \exp\left[\frac{(x_0 - x_j)^2 + (y_0 - y_j)^2}{w_0^2}\right] circ\left[\frac{\sqrt{(x_0 - x_j)^2 + (y_0 - y_j)^2}}{(d/2)}\right], \tag{1}$$

where $(x_0, y_0)$ represent the coordinates of the source plane in the *Cartesian* coordinates system, and $(x_j, y_j)$ denotes the central position of the $j$-th beamlet. $circ(\rho)$ is the transmittance function of the circular aperture, namely,

$$circ(\rho) = \begin{cases} 1 & \rho < 1 \\ 1/2 & \rho = 1 \\ 0 & otherwise \end{cases}. \tag{2}$$

Using the *Cartesian* coordinates representation, the optical field of the coherently combined beam can be accurately calculated, whereas it is difficult to illustrate the principle for the formation of the mainlobe and sidelobes of the optical field. Our motivation of this work is to extract more properties of the optical field, and the core idea is to represent the optical field of laser array by the composition of the orthogonal basis over the azimuthal plane. Therefore, prior to operating the azimuthal decomposition of the optical field, it is necessary to transform the *Cartesian* coordinates representation into the *Polar* coordinates representation by using the relations $x_0 = r_0 \cos\theta_0$ and $y_0 = r_0 \sin\theta_0$, where $(r_0, \theta_0)$ accounts for the *Polar* coordinates of the source plane. The transformed mathematical form of the laser array at the source plane yields

$$E(r_0, \theta_0) = A_0 \sum_{j=1}^{N} \exp\left[-\frac{r_0^2 + r_j^2 - 2r_0 r_j \cos(\theta_0 - \theta_j)}{w_0^2}\right] \\ \times circ\left[\frac{\sqrt{r_0^2 + r_j^2 - 2r_0 r_j \cos(\theta_0 - \theta_j)}}{(d/2)}\right]. \tag{3}$$

Taking the angular harmonics in form of $\exp(il\theta_0)$ as the basis, the expansion of the optical field at the source plane can now be expressed as

$$E(r_0, \theta_0) = \frac{1}{\sqrt{2\pi}} \sum_{l=-\infty}^{+\infty} a_l(r_0) \exp(il\theta_0), \tag{4}$$

where the complex coefficients $a_l(r_0)$ satisfy

$$a_l(r_0) = \frac{1}{\sqrt{2\pi}} \int_0^{2\pi} E(r_0, \theta_0) \exp(-il\theta_0) d\theta_0. \tag{5}$$

In the view of angular domain, angular position and angular momentum that carry the essential information of the optical field, are linked by a Fourier transformation [29,30]. The angular distribution is directly given by the complex amplitude distribution of the laser array represented in the *Polar* coordinates system. Correspondingly, the Fourier conjugate of the azimuthal angle, namely the OAM, is related to the complex coefficients of the angular harmonics. The complex amplitude of the $l$th-order OAM mode that exists in the optical field is expressed as

$$U_l(r_0, \theta_0) = \frac{1}{\sqrt{2\pi}} a_l(r_0) \exp(il\theta_0). \tag{6}$$

The relative power corresponding to the angular harmonic $\exp(il\theta_0)$ is defined as the OAM mode purity, which describes the energy proportion of the $l$th-order OAM mode that exists in the optical field. All these weights of power that correspond to the angular harmonics construct the OAM spectrum. Specifically, the purity of the $l$th-order OAM mode $P_l$ is given by [31]

$$\begin{cases} P_l = \dfrac{p_l}{\sum_{k=-\infty}^{+\infty} p_k} \\ \\ p_l = \int_0^{+\infty} |a_l(r_0)|^2 r_0 dr_0 \end{cases}. \tag{7}$$

Substituting equation (3) and equation (5) to equation (7), the OAM spectrum of the emitting laser array can be obtained.

According to the angular and OAM mode distributions determined by equations (3)-(6), the optical field of the emitting laser array at the source plane can be decomposed into two components, i.e.

$$\begin{cases} E(r_0,\theta_0) = E_c(r_0,\theta_0) + E_s(r_0,\theta_0) \\ E_c(r_0,\theta_0) = \dfrac{1}{\sqrt{2\pi}} a_0^{(0)}(r_0) \\ E_s(r_0,\theta_0) = \sum_{l=1}^{+\infty} \left[ U_l(r_0,\theta_0) + U_{-l}(r_0,\theta_0) + \dfrac{1}{\sqrt{2\pi}} a_0^{(l)}(r_0) \right] \end{cases}, \quad (8)$$

where the component $E_c(r_0, \theta_0)$ and $E_s(r_0, \theta_0)$ contribute to the formation of the mainlobe and sidelobes in the far-field, respectively. The upper index of the complex coefficient $a_0(r_0)$ denotes the diffraction order of the zeroth-order OAM mode. Hence, in the angular domain perspective of CBC, suppressing the sidelobes is synonymous with enhancing the energy percentage in the zeroth diffraction order of the zeroth-order OAM mode.

## 3. Results of angular domain insights

In CBC systems, the emitting laser array configuration at the source plane usually arranges the beamlets in terms of a centrally symmetric shape, mainly including square, hexagon, and circle. The centrally symmetric geometry of the laser array indicates that the square, hexagonal, and circular arranged coherent laser arrays can all be represented by the superposition of multiple radial subarrays and a central beamlet. Without loss of generality, a laser array of circular arrangement is studied and discussed. We firstly construct the model of circular arranged laser array and introduce its mathematical form. According to the mathematical form, the sidebands that present in the OAM spectrum of the laser array are derived, which also indicate the indices of the OAM modes that are necessary for the azimuthal decomposition of the optical field. Then, the numerical investigation of OAM spectrum is conducted to validate the theoretical prediction. Based on the results of OAM spectrum, the optical field of the laser array is azimuthally decomposed by the specific OAM modes, and the formation of mainlobe and sidelobes in the far-field are discussed. Moreover, the principle behind the engineering experience, namely fully filling the total emitting aperture with array elements to suppress the sidelobes, is elucidated in the view of angular domain.

**3.1 Theoretical analysis**

Consider an emitting laser array of circular arrangement, which contains $n$ concentrically arranged radial subarrays and an array element positioned at the center, as shown in Fig. 2(a). The $n_1$-*th* radial subarray consists of $n_1 n_0$ beamlets, and the central position of the $n_2$-*th* beamlet of the $n_1$-*th* radial subarray satisfies

$$\begin{cases} r_{n_1,n_2} = n_1 R \\ \theta_{n_1,n_2} = \dfrac{2\pi n_2}{n_0 n_1} \end{cases}, \quad (9)$$

where $R$ denotes the distance between the first (inner) radial subarray and the center of the laser array. Using equation (3) and equation (9), the optical field of the emitting laser array at the source plane can be expressed as

$$E(r_0,\theta_0) = A_0 \exp\left(-\dfrac{r_0^2}{w_0^2}\right) circ\left[\dfrac{r_0}{(d/2)}\right] \\ + \sum_{n_1=1}^{n} \sum_{n_2=1}^{n_1 n_0} \exp\left[-\dfrac{r_0^2 + n_1^2 R^2 - 2n_1 r_0 R \cos\left(\theta_0 - \dfrac{2\pi n_2}{n_0 n_1}\right)}{w_0^2}\right] \\ \times circ\left[\dfrac{\sqrt{r_0^2 + n_1^2 R^2 - 2n_1 r_0 R \cos\left(\theta_0 - \dfrac{2\pi n_2}{n_0 n_1}\right)}}{(d/2)}\right]. \quad (10)$$

Recalling the series of Bessel function [32]

$$\exp(iz\cos\theta_0) = \sum_{l=-\infty}^{+\infty} i^l J_l(z) e^{il\theta_0}, \quad (11)$$

where $J_l(\cdot)$ represents the Bessel function of the first kind and $l$th order. Therefore, we can represent the optical field of the emitting laser array by the series of angular harmonics, namely

$$E(r_0,\theta_0) = A_0 \sum_{t=1}^{N_t} B_t \left[ \exp\left(-\frac{r_0^2}{w_t^2}\right) + \sum_{l=-\infty}^{+\infty} \sum_{n_1=1}^{n} \sum_{n_2=1}^{n_1 n_0} I_l\left(\frac{2n_1 r_0 R}{w_t^2}\right) \exp\left(-\frac{r_0^2 + n_1^2 R^2}{w_t^2}\right) \right. \\ \left. \times \exp\left(-i\frac{2\pi l n_2}{n_0 n_1}\right) \exp(il\theta_0) \right], \quad (12)$$

where $I_l(z) = i^{-l} J_l(iz)$ is the modified Bessel function. Here, we have expanded the transmittance function of the circular aperture by the complex Gaussian function, i.e. [33]

$$circ\left[\frac{\rho}{(d/2)}\right] = \sum_{t=1}^{N_t} B_t \exp\left(-\frac{4C_t \rho^2}{d^2}\right), \quad (13)$$

where $B_t$ and $C_t$ are the expansion coefficients, and $t$ is the expansion order. In equation (12), The equivalent waist width $w_t$ corresponds to the order $t$ is given by

$$w_t = \frac{d}{\sqrt{4C_t^2 w_0^2 + d^2}} w_0. \quad (14)$$

Comparing equation (12) to equation (4), the simplified complex coefficients $a_l(r_0)$ of angular harmonics can now be obtained, yielding,

$$a_l(r_0) = \begin{cases} \sqrt{2\pi} A_0 \sum_{t=1}^{N_t} \sum_{n_1=0}^{n} B_t I_l\left(\frac{2n_1 r_0 R}{w_t^2}\right) \exp\left(-\frac{r_0^2 + n_1^2 R^2}{w_t^2}\right) \\ \times \left[\frac{\sin(\pi n_1)}{\pi n_1} + \frac{\sin(\pi l)}{\sin\left(\frac{\pi l}{n_1 n_0}\right)}\right] & l = mn_0, m \in Z \\ 0 & \text{otherwise} \end{cases} \quad (15)$$

Therefore, we can conclude that in the view of angular domain, the emitting laser array of circular arrangement can be decomposed by the specific angular harmonics in terms of $\exp(imn_0\theta_0)$, where $m$ is an integer. In other words, only the OAM modes of $l = mn_0$ present in the OAM spectrum of the laser array. Besides, the dominant zeroth-order OAM mode exists in the subarrays and the central beamlet, while the $n_1$-th radial subarrays contribute to the sidebands $l = \pm qn_1 n_0$ ($q \in N^+$) of the OAM spectrum.

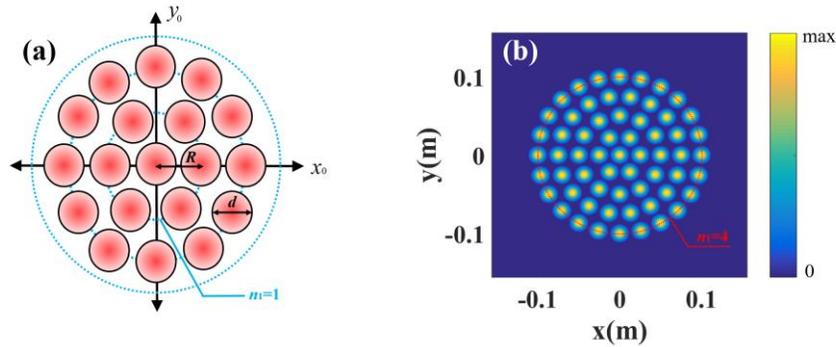

Fig. 2. Emitting laser array for the CBC system. (a) Schematic of the circular arranged laser array. (b) Intensity distribution of a 61-elements circular laser array at the source plane.

## 3.2 Numerical validation of OAM spectrum

Based on the above analysis, sidebands that present in the OAM spectrum of the laser array can be predicted. Here, numerical investigation of OAM spectrum has been carried out to validate our theoretical predictions. We take a 61-channel CBC system as an example. At the source plane, 61 circular-aperture truncated fundamental Gaussian beamlets are circularly arranged, as shown in Fig. 2(b). The parameters of the CBC system are $\lambda = 1.06$ μm, $w_0 = 10.24$ mm, $d = 23$ mm, $R = 25$ mm, $n = 4$, and $n_0 = 6$. According to our theoretical predictions, the OAM modes with the topological charges of $l = 0, \pm6, \pm12, \pm18, \ldots$ present in the OAM spectrum of the emitting laser array.

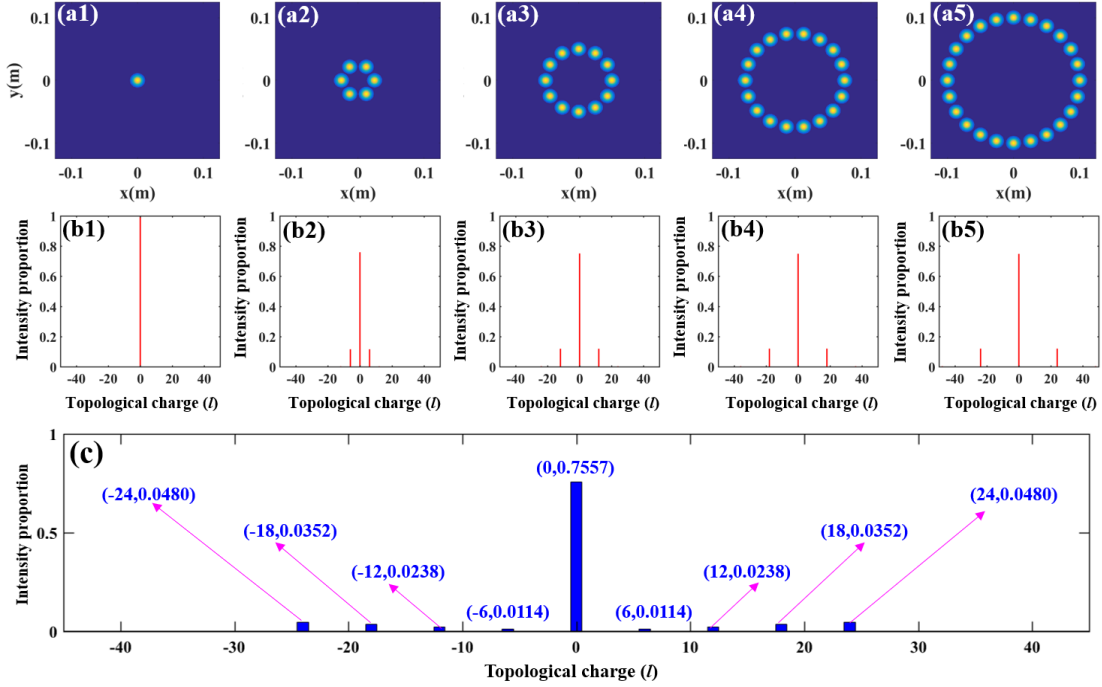

Fig. 3. Analysis on the OAM spectrum of emitting laser array. (a1)-(a5) are the intensity distributions of the central beamlet, first, second, third, and fourth radial subarrays, respectively. (b1)-(b5) are, in turns; the OAM spectra of the central beamlet, first, second, third, and fourth radial subarrays. (c) OAM spectrum of the 61-elements circular laser array.

To validate the predictions given by equation (15) of the above section, we have numerically calculated the OAM spectrum of the laser array based on the general definition shown in the *Principle* section. Figure 3 exhibits the results of OAM spectrum analysis for the 61-element coherent laser array. In our investigation, the purity of OAM modes is calculated by using equation (5) and equation (7), thus the OAM spectrum can be constructed. It is worth noting that the ranges of $l$ and $k$ in equation (5) and equation (7) are $(-\infty, +\infty)$, which indicates that the unavailable ranges should be bounded in practical analysis. Besides, when the ranges are over bounded, the accuracy of the results would be decreased. Therefore, the ranges are chosen as [-120, +120] in our study to ensure the high accuracy of calculation [34,35]. To illustrate the contribution of the central beamlet and radial subarrays to the OAM modes of the entire laser array, we firstly investigate the OAM spectra of the central beamlet, first, second, third, and fourth radial subarrays, of which the intensity distributions are shown in Figs. 3(a1)-3(a5), respectively. The corresponding OAM spectra are presented in Figs. 3(b1)-3(b5). It can be seen that the central beamlet only contains the OAM mode with a topological charge of $l = 0$. In comparison, the OAM spectra of the first, second, third, and fourth radial subarrays have main components at $l = 0$ and sidebands at $l = \pm6, \pm12, \pm18$, and $\pm24$, respectively. The purity of the OAM mode with a topological charge of $l = 0$ is around 0.75 for each radial subarray. Besides, higher-order OAM mode components exist in each subarray, while the energy proportion of the higher-order OAM mode is lower than 0.005 (for example, the purity of the OAM±48 modes in the fourth radial subarrays is 0.0039). Hence, the sidebands of these higher-order OAM modes are difficult to be observed in the OAM spectra. Figure 3(c) displays the OAM spectrum of the entire emitting laser array, which consists of the subarrays illustrated above. As a coherent superposition of the central beamlet and the radial subarrays, the 61-element laser array contains the

OAM mode with a topological charge of $l = 0$, which accounts for 0.7557 of the total energy. The OAM modes with the topological charges of -24, -18, -12, -6, +6, +12, +18, and +24 constitute the OAM sidebands, and the energy proportions are 0.0480, 0.0352, 0.0238, 0.0114, 0.0114, 0.0238, 0.0352, and 0.0480, respectively. The higher-order OAM mode components such as the OAM±36 and ±48 modes indeed exist in the laser array, as predicted theoretically. However, the energy percentage of each higher-order mode is always less than 0.002, and these sidebands are almost invisible in the presented OAM spectrum.

To conclude, the OAM spectra of the central beamlet, each radial subarray, and the entire emitting laser array calculated by numerical simulation are in consistent with the theoretical predictions, which has been derived in the above section. Based on the results of OAM spectrum analysis, we have the prior knowledge of the OAM mode components that exist in the emitting laser array, indicating that the indices of the OAM modes that are necessary for the azimuthal decomposition of the optical field are obtained, which would be further discussed in the following section.

### 3.3 Azimuthal decomposition and formation of main/side lobes

In the view of angular domain, the emitting laser array can be represented by the superposition of angular harmonics. Based on the analysis of OAM spectrum, the specific angular harmonics for optical field representation can be obtained. In other words, we have acquired the prior knowledge of the OAM mode components that exist in the emitting laser array. To further discuss the optical field of the CBC system, the spatial distributions of the specific OAM mode components at the source plane and in the far-field require to be investigated. Here, the spatial distributions of the specific OAM modes are obtained by employing the azimuthal decomposition.

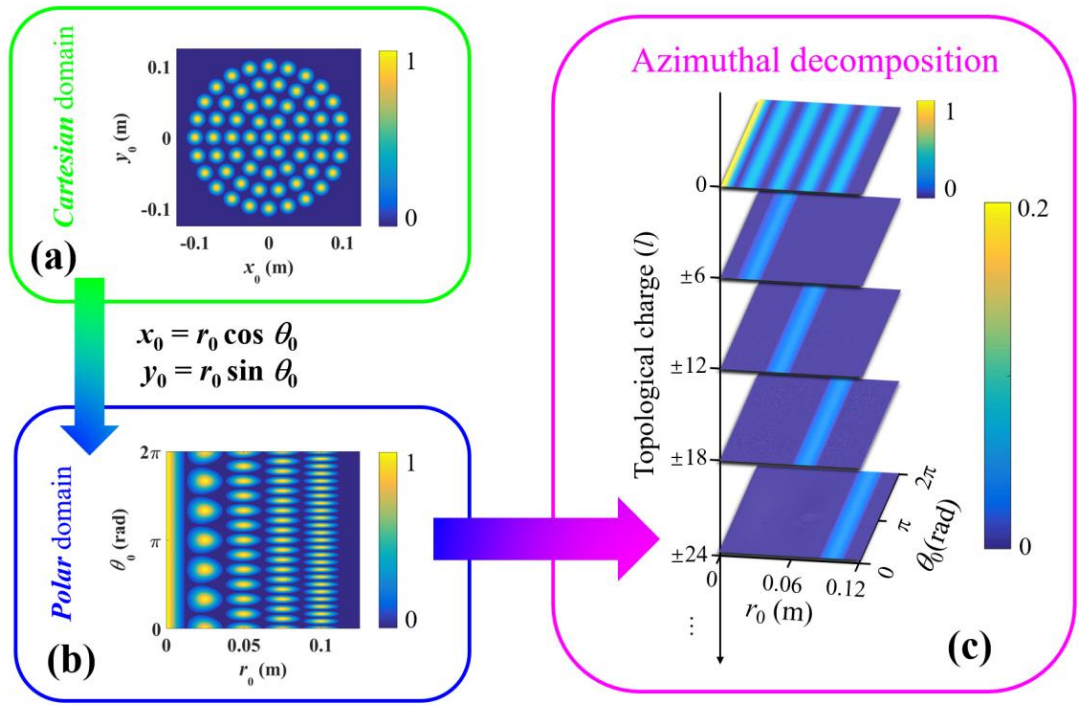

Fig. 4. Illustration of azimuthal decomposition for emitting laser array. (a) Intensity profile of the emitting laser array based on *Cartesian* coordinates representation. (b) Intensity profile of the emitting laser array based on *Polar* coordinates representation. (c) Spatial distributions of the OAM modes that exist in the emitting laser array at the source plane.

Historically, azimuthal decomposition has been celebrated for its utility in modal decomposition of laser beams [36,37]. Especially, it has been demonstrated that the azimuthal decomposition offers an accurate and robust way to decompose and reconstruct the optical field of an arbitrary laser source without any scale information, and the azimuthal decomposition of optical field can be performed by digital holograms in experiments [36]. In this work, the laser array is different from a typical solid-state laser source, which has an intrinsic scale of laser modes. Therefore, it is suitable to utilize the azimuthal decomposition method to acquire the spatial distributions of the OAM modes instead of using a set of complete and orthogonal basis with a determined intrinsic scale, such as the Laguerre-Gaussian (LG) modes and Hermite-Gaussian (HG) modes, for modal decomposition [37-39]. Besides, our motivation is to theoretically analyze the OAM modes of the emitting laser array, thus the azimuthal decomposition can be directly performed by numerical calculations. The implementation of azimuthal decomposition is schematically illustrated in Fig. 4. Initially, the *Cartesian* coordinates ($x_0$, $y_0$) have been established and the 61-element laser array is represented in the *Cartesian* coordinates system [see Fig. 4(a)]. Then, the *Cartesian* coordinates representation are transformed into the *Polar* coordinates representation by using the relations $x_0 = r_0 \cos\theta_0$ and $y_0 = r_0 \sin\theta_0$, and the intensity profile of the 61-element laser array represented in the *Polar* coordinates system is shown in Fig. 4(b). Consequently, with the prior knowledge of the OAM mode components that exist in the emitting laser array, the spatial distributions of the specific OAM modes [see Fig. 4(c)] are obtained by calculating equation (6) and equation (15). Different from the OAM spectrum that describes the energy proportions of the OAM modes in the entire optical field of the emitting laser array, the results of azimuthal decomposition provide more detailed spatial distributions of the specific OAM modes. The spatial distributions of the specific OAM modes at the source plane determine the formation of the optical field distribution of the combined beam in the far-field, which we now investigate.

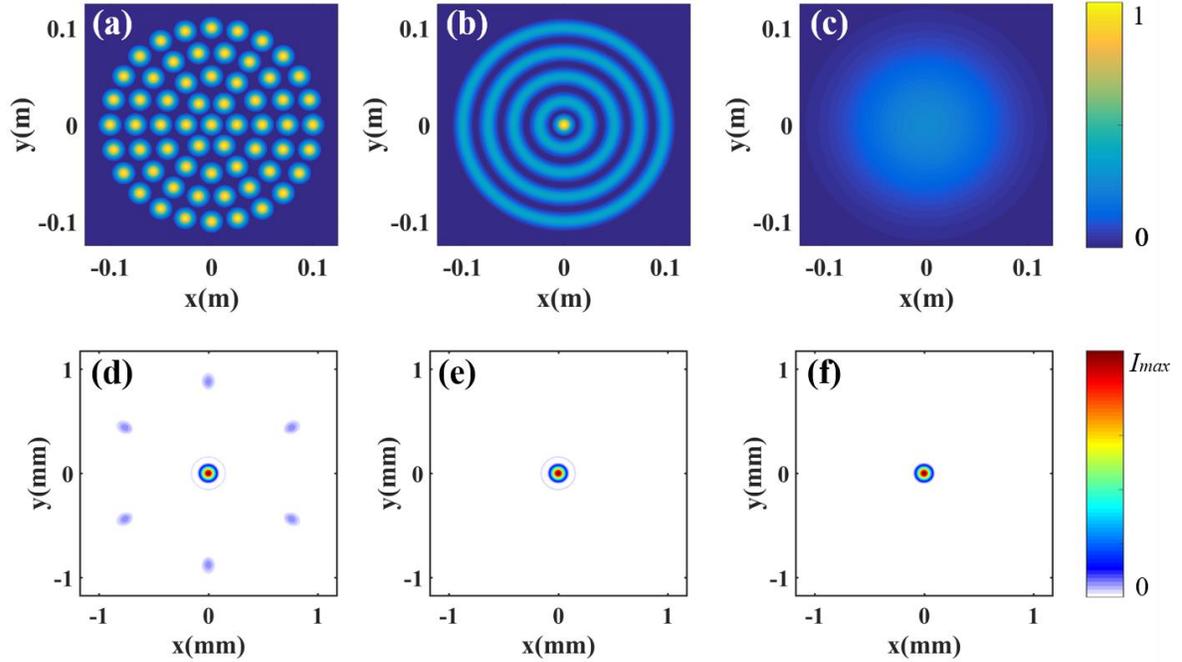

Fig. 5. Formation of mainlobe in CBC system. Intensity distributions of the (a) 61-element laser array, (b) zeroth-order OAM mode component, and (c) zeroth diffraction order component of the zeroth-order OAM mode at the source plane. (d), (e), and (f) are the far-field intensity distributions that correspond to (a), (b), and (c), respectively.

In conventional insights, the far-field intensity distribution of the CBC system is formed by the interference of beamlets with different spatial positions at the source plane of the *Cartesian* coordinates system. With the results of azimuthal decomposition shown in Fig. 4(c), the formation of far-field intensity distribution can be illustrated in the view of angular domain. Figures 5(a) and 5(d) exhibit the intensity distributions of the 61-element emitting laser array at the source plane and in the far-field, respectively. We suppose that a focus lens with a focal length of $f = 20$ m is positioned behind the laser array, thus the far-field intensity distribution of the coherently combined beam can be observed at the focal plane. The far-field intensity distribution shown in Fig. 5(d) has a mainlobe of concentrated energy and sidelobes surrounding the mainlobe. The intensity distribution of the zeroth-order OAM mode that exists in the 61-element laser array at the source plane is displayed in Fig. 5(b). To facilitate comparison with the intensity distribution of the entire laser array, the intensity distribution of the zeroth-order OAM mode calculated in the *Polar* domain is now represented in the *Cartesian* coordinates system. It can be seen that the nonzero area of the zeroth-order OAM mode component is matched with the central beamlet and the radial subarrays of the 61-element laser array, while only the intensity distributions in the area of the central beamlet are entirely the same. This phenomenon is consistent with the theoretical analysis in *Theoretical analysis* section, namely all the radial subarrays and the central beamlet contribute to the OAM mode of $l = 0$. The intensity distribution of the zeroth-order OAM mode in the far-field is shown in Fig. 5(e). One can see that the OAM mode of $l = 0$ that exists in the emitting laser array at the source plane would not only form the mainlobe of the coherently combined beam, but also form the surrounding concentric side rings in the far-field. These concentric side rings are the higher-order diffraction fringes of the zeroth-order OAM mode. The zeroth-order diffraction fringe of the zeroth-order OAM mode denotes the mainlobe of the coherently combined beam, which is shown in Fig. 5(f). As for the laser array of circular arrangement, the mainlobe has a Gaussian-like profile, and the corresponding intensity distribution at the source plane is of Gaussian-like profile as well. Figure 5(c) depicts the Gaussian-like intensity profile of the zeroth diffraction order of the zeroth-order OAM mode at the source plane, of which the mathematical form is expressed as $(2\pi)^{-1} |a_0^{(0)}(r_0)|^2$. Therefore, we can conclude that the zeroth diffraction order component of the OAM mode of $l = 0$ contributes to the formation of mainlobe in the far-field.

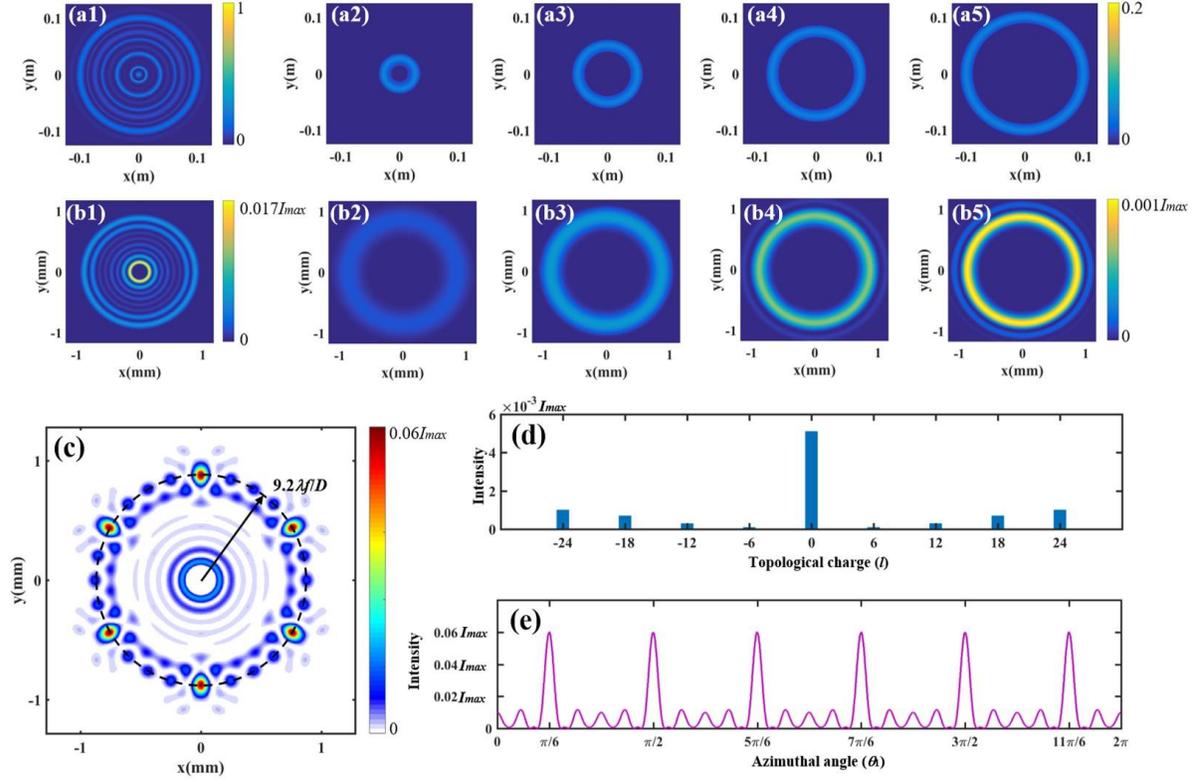

Fig. 6. Formation of sidelobes in CBC system. (a1) Intensity distribution of higher diffraction order components of the zeroth-order OAM mode at the source plane. Intensity distributions of the OAM mode components with (a2) $l = \pm 6$, (a3) $l = \pm 12$, (a4) $l = \pm 18$, and (a5) $l = \pm 24$ at the source plane. (b1)-(b5) are the intensity distributions in the far-field that correspond to (a1)-(a5), respectively. (c) Intensity distribution of the sidelobes. (d) and (e) are the OAM spectrum and intensity profile of the sidelobes in the azimuthal direction (at the radial position of $9.2\lambda f/D$, where $D = 2nR + d$), respectively.

Furthermore, the formation of sidelobes in the CBC system deserves discussion. In the above discussion of the mainlobe, we have illustrated that the OAM mode of $l = 0$ at the source plane consists of the zeroth diffraction order component and higher-order diffraction components. Figure 6(a1) exhibits the intensity distribution of higher-order diffraction components of the zeroth-order OAM mode at the source plane. Based on the results of azimuthal decomposition, the intensity distributions of the OAM mode components with $l = \pm 6$, $l = \pm 12$, $l = \pm 18$, and $l = \pm 24$ at the source plane are presented in Figs. 6(a2)-6(a5), respectively. According to the paraxial diffraction theory, the intensity distributions in the far-field correspond to Figs. 6(a1)-6(a5) are calculated and shown in Figs. 6(b1)-6(b5), respectively. The higher-order diffraction fringes of the zeroth-order OAM mode shown in Fig. 6(b1) are the above-mentioned concentric side rings, while the intensity is significantly lower than the intensity peak of the mainlobe (zeroth-order diffraction fringe). The spatial distributions of the higher-order OAM mode components are matched with the radial subarrays, which have been theoretically predicted in *Theoretical analysis* section. Specifically, the OAM mode component with the topological charge of a larger absolute value mainly occurs at the radial subarray with a larger radius, indicating that the scales of the spatial modes with the different $|l|$ are different at the source plane. Meanwhile, in terms of the propagating properties of optical vortex fields with the helical phase of the mathematical form $\exp(il\theta_0)$, the propagating fields of the $l$th-order OAM modes have a radial profile whose radius scales with the $l$ index [40,41]. Consequently, the sizes of the far-field intensity distributions of the OAM modes with different orders are approximately the same, and the intensity peak of each OAM mode locates around the radial position of $9.2\lambda f/D$ in the far-field, where $D = 2nR + d$ represents the total emitting diameter of the laser array. In specific, the total emitting diameter of the 61-element laser array in our study is calculated as $D = 223$ mm.

In the far-field, the optical fields of the nonzeroth-order diffraction fringes of the zeroth-order OAM mode, and the higher-order OAM modes are coherently overlapped and form the sidelobes of the combined beam, and the intensity profile of the sidelobes is shown in Fig. 6(c). It can be seen that the sidelobes of the combined beam contain concentric side rings surrounding the mainlobe and multiple petals away from the center. In the area where the side rings are formed, the intensity of the higher-order OAM modes is approximately zero due to the helical phase structures that carry optical vortex cores. Hence, the side rings close to the center of the combined beam are in consistent with the first several diffraction fringes of the zeroth-order OAM mode depicted in Fig. 6(b1). In comparison, the petals away from the center are formed by the coaxial interference of the higher-order OAM modes and higher-order diffraction fringes of the zeroth-order OAM mode. Among these petals, there are six petals uniformly distributed along the azimuthal direction with significantly higher power density, which can also be observed in Fig. 5(d). To elucidate the reason behind the formation of the six obvious petals, the localized OAM spectrum and intensity profile in the azimuthal direction (at the radial position of $9.2\lambda f/D$) have been explored, as shown in Figs. 6(d) and 6(e), respectively. Analogous to the Fourier relationship between time and frequency signals, the OAM spectrum and azimuthal intensity distribution are linked by the Fourier relationship in the view of angular domain [29,30,42]. The OAM spectrum is constructed by using the combination of multiple weighted OAM modes. When the indices of OAM modes are $l = 0, \pm 6, \pm 12, \pm 18$, and $\pm 24$, the OAM spectrum is sampled at the interval of 6. Hence, the azimuthal intensity forms the replicas in the spatial domain with a periodic of $\pi/3$ according to the Fourier relationship between the angular position and OAM, and the intensity peak of the six obvious petals locates at the angular position of $\pi/6, \pi/2, 5\pi/6, 7\pi/6, 3\pi/2$, and $11\pi/6$, respectively.

In a nutshell, we have discussed the OAM modes of the emitting laser array that contribute to the formation of mainlobe and sidelobes in the far-field based on the results of azimuthal decomposition. The OAM spectrum and spatial distributions of OAM modes are determined by the emitting laser array configuration of the CBC system at the source plane. In the far-field, the mainlobe of the combined beam is the zeroth diffraction order component of the $l = 0$ OAM mode, while the sidelobes are formed by the coaxial combination of the higher-order OAM modes and nonzeroth-order diffraction fringes of the zeroth-order OAM mode. In the angular domain perspective, the enhancement of combining efficiency (energy percentage in the mainlobe) can be considered as the optimization of the zeroth diffraction order component of the $l = 0$ OAM mode carried by the emitting laser array.

### 3.4 Illustration of engineering experience

In the view of angular domain, the formation of coherently combined beam can be considered as the coaxial superposition of the OAM modes that exist in the emitting laser array. The weights and spatial distributions of OAM modes are determined by the arrangement of the laser array at the source plane. With these insights of the emitting laser array configuration in the angular domain, we now return to the generally recognized engineering experience of the CBC systems, namely enhancing the filling of the laser array in the near field to improve the combining efficiency [1,2].

To characterize the filling of emitting laser arrays, the concept of fill factor is widely utilized. In fact, the definitions of fill factor have been proposed diversely in previous publications, which includes the conformal/subaperture fill factor [21], lenslet fill factor [43], Gaussian fill factor [24], and areal fill factor [2]. Despite the difference in definitions, these defined fill factors are appropriate to reflect the filling of the laser arrays. In this work, we take the areal fill factor $F = Nd^2 / D$ as an example, which is defined as dividing the area of the tiled apertures by the total area of the total emitting aperture. In practical implementations of CBC, there is a conventional experience namely engineering the maximum areal fill factor as possible. Here, the engineering experience is elucidated in the view of angular domain.

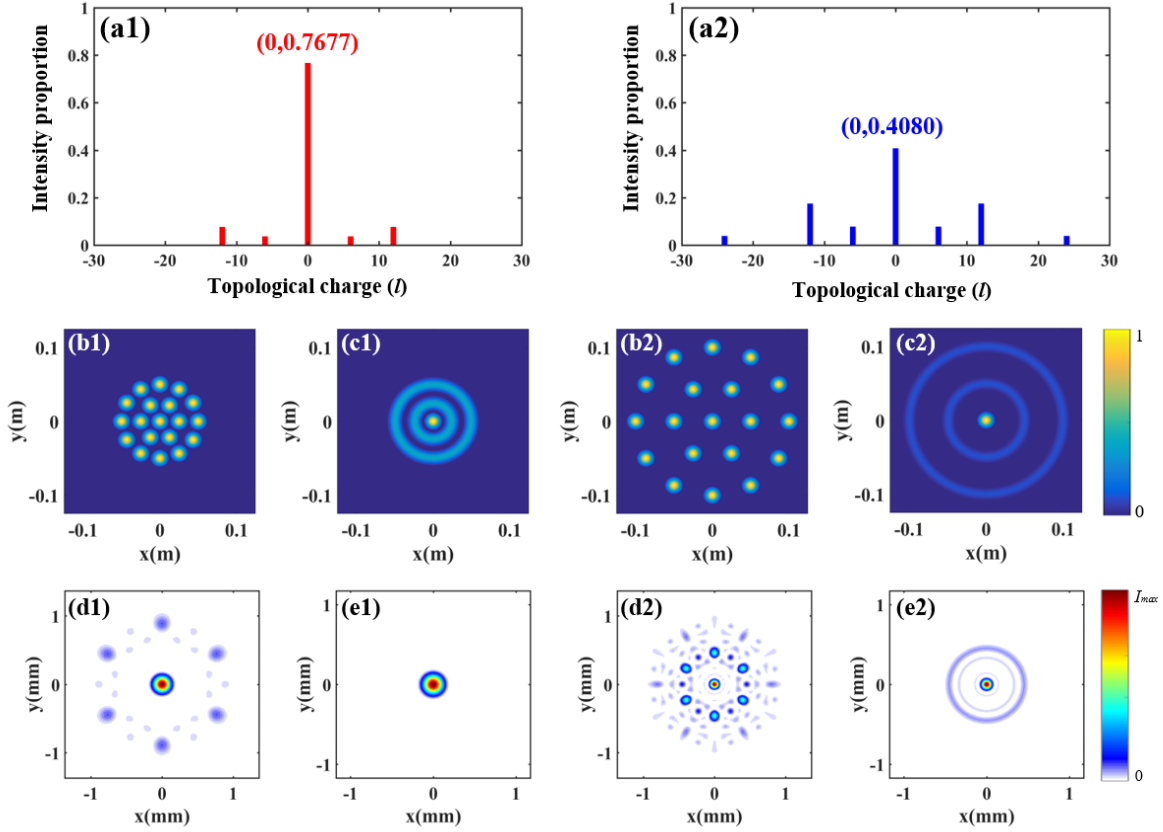

Fig. 7. Comparison of emitting laser arrays with different areal fill factors. OAM spectra of the 19-element laser array with the fill factors (a1) $F = 0.7$ and (a2) $F = 0.2$. (b1) Intensity distribution and (c1) zeroth-order OAM mode of the laser array with $F = 0.7$ at the source plane. (b2) Intensity distribution and (c2) zeroth-order OAM mode of the laser array with $F = 0.2$ at the source plane. (d1), (e1), (d2), and (e2) are the far-field intensity distributions correspond to (b1), (c1), (b2), and (c2), respectively.

Without loss of generality, 19-element circular arranged emitting laser arrays with different fill factors are taken as an example. The wavelength, waist width, and diameter of each array element are set to be $\lambda = 1.06$ μm, $w_0 = 10.24$ mm, and $d = 23$ mm, respectively. As for the compactly arranged laser array, the distance between the first (inner) radial subarray and the center $R = 25$ mm. Accordingly, the diameter of total emitting aperture $D = 123$ mm, and the areal fill factor is calculated as $F = 0.664$. When the parameter $R$ is set to be 50mm, the diameter of total emitting aperture $D$ is enlarged to 223 mm, and the areal fill factor $F$ is correspondingly decreased to 0.202. The comparison of emitting laser arrays with the areal fill factors $F = 0.664$ and $F = 0.202$ has been conducted and the results are shown in Fig. 7. Figures 7(a1) and 7(a2) display the OAM spectra of the laser arrays with the areal fill factors of 0.664 and 0.202, respectively. It can be observed that when the spacing between the adjacent beamlets increases, the energy of the main OAM mode component spreads to the sidebands, and the purity of the zeroth-order OAM mode greatly decreases from 0.7677 to 0.4080. The higher-order OAM mode components such as the OAM±24 modes not only exist in the radial subarray, but are obvious in the presented OAM spectrum as well. The intensity distributions of the laser array with $F = 0.664$ and the corresponding zeroth-order OAM mode at the source plane are shown in Figs. 7(b1) and 7(c1), respectively, which are similar with the case of the 61-element compactly arranged laser array. In the far-field, the intensity distribution of the combined beam has a mainlobe of concentrated energy and surrounding sidelobes [see Fig. 7(d1)]. Besides, the existed zeroth-order OAM mode evolves through propagation and forms the diffraction fringes in the far-field, of which the intensity distribution is depicted in Fig. 7(e1). On the contrary, the intensity distributions of the laser array with $F = 0.202$ and the existed zeroth-order OAM mode at the source plane, and the far-field intensity distributions of the combined beam and the zeroth-order OAM mode are presented in Figs. 7(b2), 7(c2), 7(d2), and 7(e2), respectively. The results indicate that the energy proportion of the OAM mode component with $l = 0$ decreases with the reduction of the areal fill factor, and the spatial distribution of the zeroth-order OAM mode at the source plane is correspondingly changed as well. For one thing, the increased discreteness of the optical field along azimuthal direction at the source plane leads to the reduction of the energy percentage in the zeroth-order OAM mode, thus the total energy of the far-field diffraction fringes of the zeroth-order OAM mode is decreased. For another thing, the discreteness of the optical field along radial direction at the source plane is also increased, which causes the redistributions of the zeroth-order OAM mode at the source plane, and accordingly, the energy of the zeroth diffraction order of the zeroth-order OAM mode in the far-field spreads to the higher-order diffraction fringes.

To summarize, the engineering experience related to the fill factor has been illustrated in the angular domain perspective. With the reduction of the areal fill factor, the discreteness of the optical field along both the azimuthal and radial directions at the source plane increases, which would cause the energy spreading of the OAM spectrum and the energy spreading of the diffraction fringes of the zeroth-order OAM mode in the far-field, respectively. As a result, the energy spreads from the mainlobe to the surrounding sidelobes in the far-field, and the combining efficiency would be degraded. Hence, the emitting laser array configuration of the CBC system should be designed of a close-packed arrangement, and the areal fill factor is engineered to be close to 1.

## 4. Fractal-based optimization

According to the theoretical analysis and results of the emitting laser arrays in the angular domain perspective, we have found that the mainlobe of the coherently combined beam is synonymous with the zeroth-order diffraction fringe of the OAM mode component with $l = 0$. To suppress the sidelobes and enhance the combining efficiency, on the one hand, the OAM spectrum of the laser array can be optimized by reducing the discreteness of the optical field along azimuthal direction at the source plane, and on the other hand, the far-field diffraction fringes of the zeroth-order OAM mode can be optimized by structuring the spatial distribution of the zeroth-order OAM mode at the source plane. Here we propose a design of emitting laser array configuration based on the self-similar fractal structure to realize the optimization of the spectrum and spatial distribution of OAM modes.

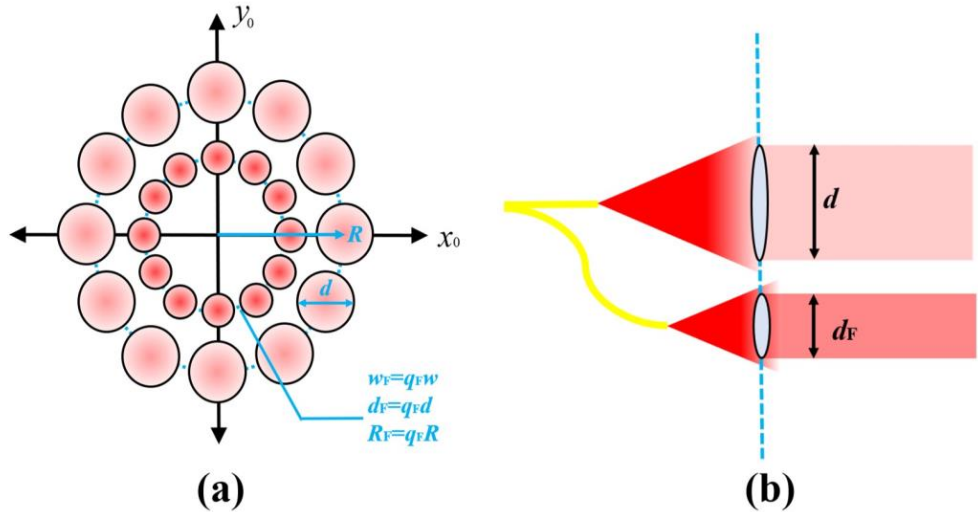

Fig. 8. The sketch for the fractal-based configuration of emitting laser array. (a) Arrangement of the fractal laser array in the $x_0y_0$-plane. (b) Setup of the fractal laser array in the $y_0z_0$-plane.

Fractals are complex shapes with self-similarity, which have been observed widely in nature [44]. For a long time, the marriage between fractals and light has incubated striking achievements [45-49]. Especially, the controlled generation of fractal light inside a single laser cavity has been experimentally demonstrated in recent years [48]. As for the spatially distributed laser arrays in CBC systems, the introduction of fractals is expected to bring fascinating properties as well. Figure 8(a) schematically depicts the arrangement of a typical fractal-based emitting laser array that lies in the $x_0y_0$-plane. The inner and outer radial subarrays that contains the beamlets of the same number constitute a set of simple fractal configuration with a scaling factor of $q_F$. When the waist width and aperture diameter of each array element lying on the outer radial subarray, and the radius of the outer subarray are $w$, $d$, and $R$ respectively, the corresponding parameters of the self-similar inner subarray are set to be $w_F = q_Fw$, $d_F = q_Fd$, and $R_F = q_FR$, respectively. It is worth noting that when the CFAs of each channel operate at the maximum output power, the amplitude of the beamlet lying on the outer subarray is $q_F$ times the amplitude of the beamlet lying on the inner subarray, according to the conservation of energy. Therefore, the fractal configuration can tailor the dimension, position, and geometric arrangement of the elements on the laser array and naturally introduce the amplitude modulation of beamlets at the same time. To realize the fractal configuration in practical implementations, the setup the fiber and collimator arrays can be designed in terms of Fig. 8(b). When the parameters of the fibers and fractal-based coherent laser array are determined, the distance between the fiber tail and the collimator can be derived and calculated. In CBC systems, the fractal configuration presented in Fig. 8 is compatible with the conventional arrangement of beamlets, and the appropriate use of fractal configuration in emitting laser arrays is capable of optimizing the spectrum and spatial distributions of OAM modes.

To demonstrate the utility of fractal-based optimization, investigations on the performance of the conventional laser array and the optimized laser array that employs the fractal configurations have been carried out. For comparison, we assume that both the conventional and optimized laser arrays consist of 37 channels laser beams and synthesize a total emitting aperture with a diameter of 173 mm. For each channel, the CFAs operate at the maximum output power. In the 37-element conventional laser array that contains a central beamlet and three radial subarrays, the waist width, and aperture diameter of each array element are set to be $w_0 = 10.24$ mm and $d = 23$ mm, respectively, and the radius of the first (inner) radial subarray is $R = 25$ mm. As for the emitting laser array with fractal-based optimization, a beamlet is positioned at the center, while the other 36 beamlets are arranged in four radial subarrays. The first and second subarrays constitute a set of fractal configuration with the scaling factor of 0.37, and the third and fourth subarrays are also constructed in fractal configuration with the scaling factor of 0.63. Specific parameters of the optimized laser array that employs fractal configurations are listed in Table 1.

Table 1  Parameters of the emitting laser array with fractal-based optimization

|  | Central beamlet | Fractal configuration 1 | | Fractal configuration 2 | |
| --- | --- | --- | --- | --- | --- |
| Serial number of subarrays | -- | 1 | 2 | 3 | 4 |
| Waist width (mm) | 3.3 | 3.3 | 9.0 | 9.0 | 14.4 |
| Aperture diameter (mm) | 7.5 | 7.5 | 20.3 | 20.3 | 32.3 |
| Radius of subarray (mm) | -- | 8.1 | 22.0 | 44.0 | 70.3 |
| Normalized amplitude | 1 | 1 | 0.37 | 0.37 | 0.23 |
| Number of beamlets | 1 | 6 | 6 | 12 | 12 |
| Scaling factor | -- | 0.37 | | 0.63 | |

At the source plane, the arrangement and parameters of the emitting laser array determines the OAM spectrum and the spatial distributions of the existed OAM modes. Figure 9 exhibits the comparison of the conventional laser array and the optimized laser array at the source plane. The intensity distributions of the 37-element conventional laser array and the existed zeroth-order OAM mode at the source plane are shown in Figs. 9(a) and 9(b), respectively. In comparison, the intensity distributions of the laser array with fractal-based optimization and the corresponding zeroth-order OAM mode are displayed in Figs. 9(c) and 9(d), respectively. Since the output power of the CFAs is the same for each channel, the intensity profiles of the beamlets lying on the conventional laser array are the same, whereas the amplitude of the beamlet lying on the optimized laser array varies with the dimension of the waist width and aperture. In fractal configurations, the inner subarray has a smaller dimension compared to the outer subarray, thus the amplitudes of the beamlets lying on the outer subarray are always lower. In the prior study, active amplitude modulation has been utilized to tailor the beamlets of laser array for fitting a Gaussian envelope, which can efficiently enhance the energy concentration of the far-field [50]. Here, the fractal configurations we proposed naturally introduce the amplitude modulation, indicating that the effect of apodization can be achieved without the loss of total energy. The OAM spectra of the conventional and optimized laser array are presented in Figs. 9(e) and 9(f), respectively. According to the azimuthal distributions of the array elements, the laser array with fractal-based optimization has secondary OAM modes at $l = \pm 12$, and the weights are both 8.3%, whereas the dominant OAM sidebands of the conventional laser array are $l = \pm 18$ with the same weights of 5.8%. It can be observed that the use of fractal-based optimization can improve the energy proportion of the main OAM mode component ($l = 0$) from 75.6% to 77.8%.

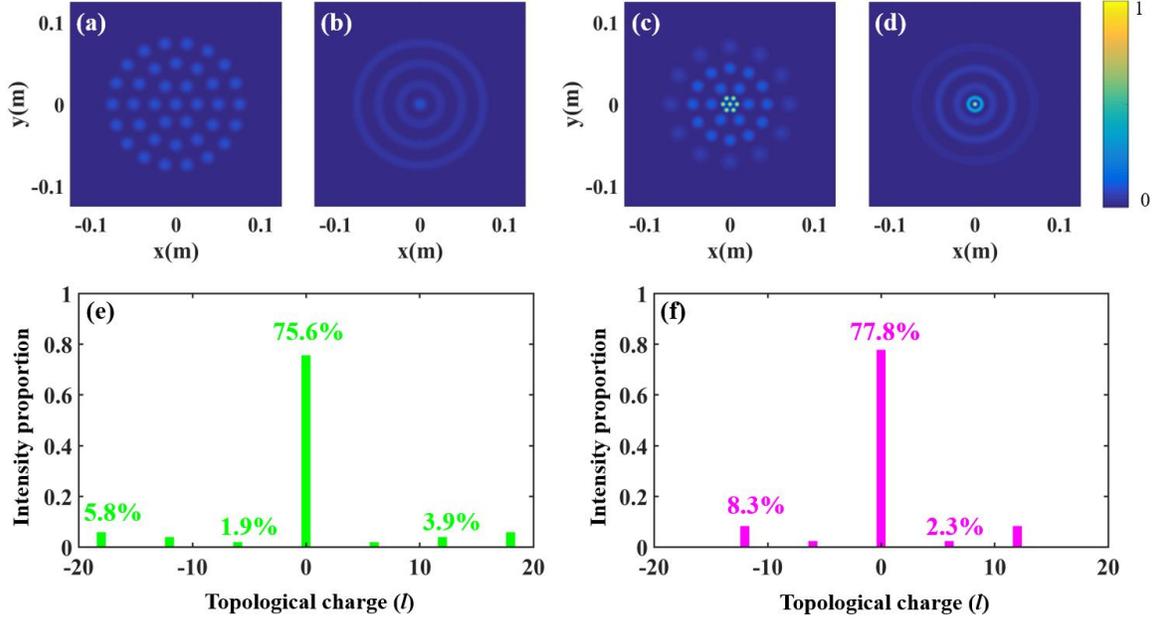

Fig. 9. Comparison of the conventional laser array and the laser array with fractal-based optimization at the source plane. (a) Intensity distribution and (b) zeroth-order OAM mode distribution of the 37-element conventional laser array at the source plane. (c) Intensity distribution and (d) zeroth-order OAM mode distribution of the 37-element laser array with fractal-based optimization, at the source plane. (e) and (f) are the OAM spectra of the conventional laser array and the laser array with fractal-based optimization, respectively.

Furthermore, the features and performance of the coherently combined beams in the far-field have been studied for the cases of conventional laser array and the laser array with fractal-based optimization. Figures 10(a) and 10(b) show the far-field intensity distributions of the zeroth-order OAM modes corresponding to the conventional laser array and the optimized laser array, respectively. For the zeroth-order OAM mode that exists in the conventional laser array, the secondary diffraction fringe in the far-field can be observed. On the contrary, the zeroth-order OAM mode of the laser array with fractal-based optimization forms the zeroth-order diffraction fringe of concentrated energy in the far-field owing to the spatial structuring of the laser array at the source plane. The difference between the intensity distributions of the coherently combined beams that are generated from the conventional laser array and the optimized laser array is obvious, which are displayed in Figs. 10(c) and 10(d), respectively. The interference of the OAM modes with $l = 0, \pm 6, \pm 12$, and $\pm 18$ that exist in the conventional laser array forms the six azimuthally distributed petals, which are similar with the above-mentioned 19-element and 61-element laser arrays cases. By employing the fractal configurations, the OAM mode components with $l = \pm 18$ are reduced, and therefore, the optimized laser array forms the coherently combined beam with twelve azimuthally distributed petals that are of lower energy and closer to the mainlobe.

In CBC systems, beam propagation factor (BPF) metric is widely used to evaluate the performance of the coherently combined beams [23,50,51]. The definition of BPF metric is closely related to the power in the bucket (PIB) [17,21], which describes the power encircled in an on-axis circular area with a specific radius (bucket radius) at the focal plane ($z = f$). Specifically, the BPF metric is defined as the output power of the combined beam encircled in the far-field bucket with the radius of $1.22\lambda f/D$, namely the PIB with the bucket radius of $1.22\lambda f/D$, divided by the effective output power radiating from the laser array at the source plane. According to the definition, the BPF metric can be expressed as

$$BPF = \frac{\int_0^{\frac{1.22\lambda f}{D}} \int_0^{2\pi} I_{combined}(\rho,\varphi) \rho d\varphi d\rho}{\int_0^{\infty} \int_0^{2\pi} \left[ E(r_0,\theta_0) E^*(r_0,\theta_0) \right] r_0 d\theta_0 dr_0} ,  \quad (16)$$

where $I_{combined}(\rho, \varphi)$ represents the far-field intensity distribution of the combined OAM beam, and ($\rho, \varphi$) accounts for the *Polar* coordinates of the focal plane.

To investigate the performance of the combined beam, the normalized PIB of the combined beam at the focal plane as a function of the bucket radius is calculated, as shown in Fig. 10(e). The diameter of the total emitting aperture is 173 mm, and the focal length is set as 20 m, thus the bucket radius related to the definition of BPF metric is calculated as 0.15 mm. It can be observed that when the bucket radius is below 0.1 mm, the PIB curves for the conventional (green) and optimized (purple) laser arrays almost coincide. With the increase of the bucket radius, the PIB of the optimized laser array is always higher when compared to the conventional laser array until the area of the bucket is close to the six petals of the combined beam of the conventional laser array. Then, the PIB for the two cases gradually approaches 1 as the bucket radius further increases. To be more specific, we take a closer look at the PIB curves around the bucket radius for calculating BPF metric, that is, 0.15 mm. Figure 10(f) depicts the locally magnified PIB curves corresponding to the ranges bounded by the blue dashed rectangle in Fig. 10(e). The advantage of employing fractal configurations can be observed, and the PIB with the bucket radius of 0.15 mm for the cases of conventional and optimized laser arrays are 0.537 and 0.491, respectively, indicating that by using the fractal-based optimization, the BPF can be improved by 9.4%.

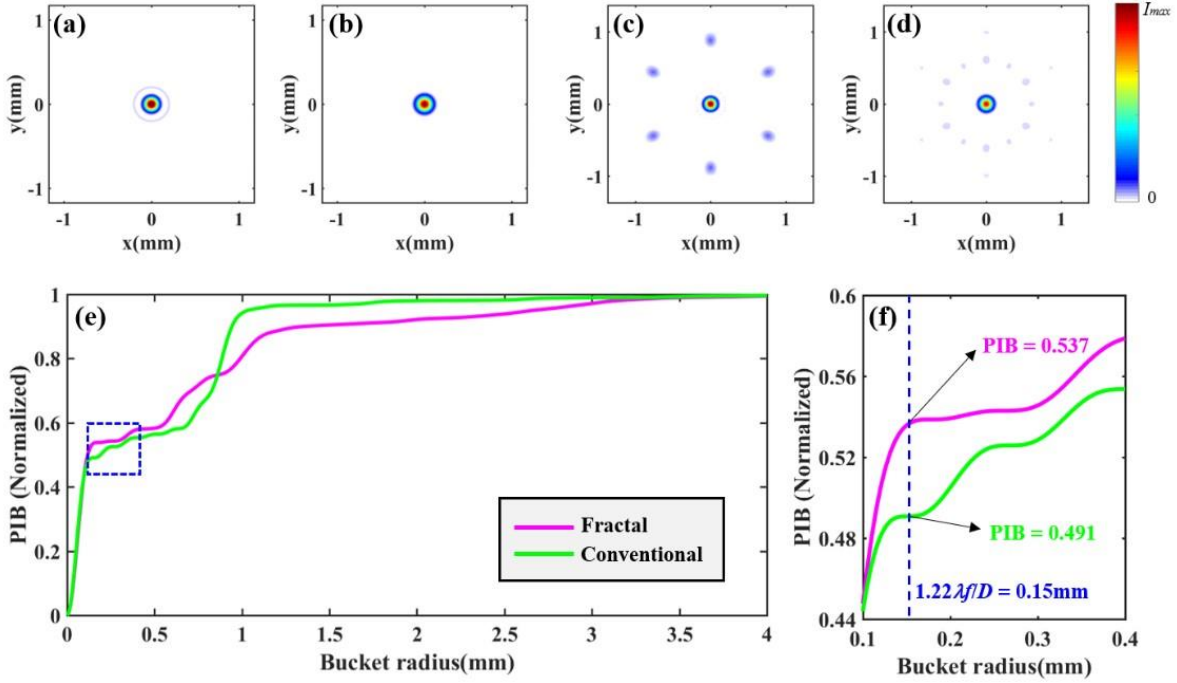

Fig. 10. Comparison of the coherently combined beams in the far-field generated by the conventional and optimized laser arrays. Far-field intensity distributions of the zeroth-order OAM modes in (a) the conventional laser array and (b) laser array with fractal-based optimization. Far-field intensity distributions of the combined beams generated from the (c) conventional laser array and (d) laser array with fractal-based optimization. (e) Power in the bucket (PIB) as a function of the bucket radius for the coherently combined beams that correspond to the conventional laser array (green) and laser array with fractal-based optimization (purple). (f) Locally magnified PIB curves for the cases of conventional laser array and optimized laser array.

In this section, we propose a fractal-based optimization method for the emitting laser array configuration of CBC systems. By employing the fractal configurations, the main OAM mode component of the OAM spectrum and the energy percentage in the zeroth-order diffraction fringe of the zeroth-order OAM mode can be efficiently enhanced. As a result, the improvement in the performance of the combined beam by using the fractal-based optimization has been demonstrated. In practical implementations, the fractal configurations we proposed are compatible with the conventional emitting laser array, and the self-similar feature can be extended to the emitting laser array of square or hexagonal arrangement. The appropriate use of fractal configurations could provide reference on the further enhancement in the performance of CBC systems.

## 5. Conclusions

In this work, we have proposed the principle and method to study the emitting laser array of CBC systems in the view of angular domain. The OAM mode components and the corresponding power proportions that construct the OAM spectrum of the emitting laser array are determined through the angular harmonic representation of the laser array, and the spatial distributions of the existed OAM modes in the laser array at the source plane can be obtained by using azimuthal decomposition. The OAM modes of the emitting laser array at the source plane propagate and coherently overlap to form the combined beam with a mainlobe and surrounding sidelobes in the far-field. We have found that the zeroth-order diffraction fringe of the OAM mode with $l = 0$ denotes the mainlobe of the combined beam, while the surrounding sidelobes are formed by the coaxial interference of the higher-order OAM modes and nonzeroth-order diffraction fringes of the zeroth-order OAM mode.

Up to this point, we have illustrated that in the view of angular domain, the power enhancement in the mainlobe of the coherently combined beam requires the optimization of zeroth diffraction order component of the zeroth-order OAM mode carried by the laser array. The optimization of the zeroth diffraction order component of the zeroth-order OAM mode depends on the purity improvement of the zeroth-order OAM mode in the entire laser array on the one hand, and the suppression of energy spreading in the diffraction fringes of the zeroth-order OAM mode in the far-field on the other hand. The optimization in terms of these two aspects can be achieved by a long-standing and intuitive engineering experience, namely maximizing the areal fill factor and limiting the discreteness of the optical field at the source plane. Moreover, a fractal-based optimization method that is compatible with the conventional emitting laser array configuration of CBC systems has been proposed, which can further improve the OAM mode purity and the energy proportion in the zeroth-order diffraction fringe of the zeroth-order OAM mode. The utility of the fractal-based optimization method has been demonstrated and the salient feature of the fractal configuration, i.e., natural amplitude modulation, could also be beneficial for spatial light structuring.

In conclusion, our work gives a starting point to study laser arrays combination in the angular domain perspective. We believe that the more detailed optical field information of CBC systems that emerges in the angular domain can not only guide the performance improvement of the coherently combined beam, but also offers exciting avenues for the structured light fields customization and phase control system upgradation, which deserve further exploration.

# Acknowledgements

We are grateful for financial supports from National Natural Science Foundation of China (NSFC) (Grant No. 62075242), Hunan Provincial Innovation Construct Project (Grant No. 2019RS3017), and the Natural Science Foundation of Hunan province, China (Grant No. 2019JJ10005). We sincerely appreciate Zongfu Jiang, Wei Liu, and Xiaoming Xi for valuable discussions.